# Future Trends in Linacs

*A. Degiovanni*
CERN, Geneva, Switzerland

**Abstract**
High-frequency hadron-therapy linacs have been studied for the last 20 years and are now being built for dedicated proton-therapy centres. The main reason for using high-frequency linacs, in spite of the small apertures and low-duty cycle, is the fact that, for such applications, beam currents of the order of a few nA and energies of about 200 MeV are sufficient. One of the main advantages of linacs, pulsing at 200–400 Hz, is that the output energy can be continuously varied, pulse-by-pulse, and a moving tumour target can be covered about ten times in 2–3 minutes by deposing the dose in many thousands of 'spots'. Starting from the first proposal and the on-going projects related to linacs for medical applications, a discussion of the trend of this field is presented focussing, in particular, on the main challenges for the future, such as the reduction of the footprint of compact 'single-room' proton machines and the power efficiency of dual proton and carbon-ion 'multi-room' facilities.

## 1   Introduction

During this lecture, a short overview of linac technology for medical applications is given, with an emphasis on existing and future projects. In particular, the case of linear accelerators for hadron therapy is discussed.

The use of protons and hadrons for the therapy of deep-seated tumours was first proposed in 1946 by Wilson in a famous paper [1]. Since then, hadron therapy has developed in the last 70 years as an advanced technique in radiation therapy, allowing non-invasive and precise irradiation of solid tumours, with the advantage of sparing the surrounding healthy tissues. This is due to the presence of the Bragg peak in the depth-dose profile of charged hadrons, compared to the exponential decay of the dose with depth typical of X-rays (Fig. 1). The overlap of many Bragg peaks, obtained by adjusting the proton beam energy, allows the production of a flat dose distribution over the depth of the tumours, the so-called Spread-Out Bragg Peak (SOBP) shown in Fig. 1. The integral dose delivered to the surrounding healthy tissues with protons and charged hadrons is 3–4 times smaller than with X-rays.

In the past decades, more than 100,000 patients have been treated with proton beams, and more than 10,000 with carbon ions. Other species, such as helium, oxygen or neon, have been used in a reduced number of cases. The typical energies used for the treatment of tumours seated at a maximum depth of 27 cm are 200 MeV for protons and 400 MeV/u for fully stripped carbon ions. In terms of beam currents, typical values are 1 nA for protons and 0.1–0.2 nA for carbon ions, corresponding to a typical dose rate of 2 Gy/l in one minute.

X-rays have two main problems in the treatment of deep seated tumours: (i) they deposit unwanted dose in the critical organs close to the target volume (see Fig. 1) and (ii) they cannot cure the so-called 'radio-resistant' tumours (about 5% of the total) that are less sensitive to radiation than the surrounding normal tissues.

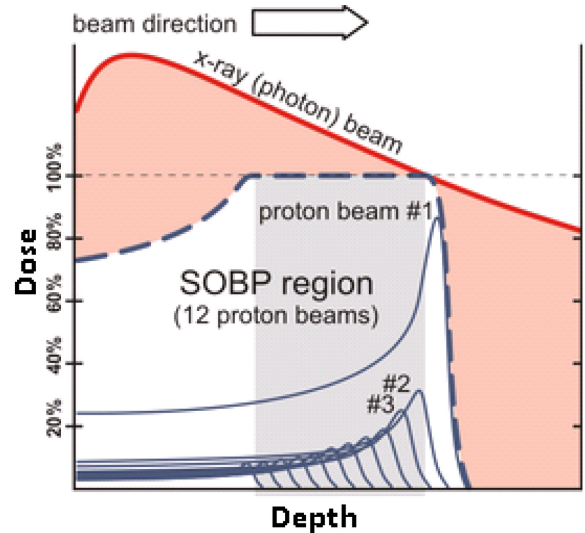

**Fig. 1:** Depth-dose profiles for X-rays and proton beams

In this respect, protons and charged hadrons, in general, can improve the treatment outcome: first of all, because they spare normal tissues, due to the lower dose delivered in the entrance channel and the practically zero dose delivered in the exit channel (a very small amount of dose is still present due to nuclear fragmentation effects of the treatment beam in the body of the patient); and secondly, because carbon ions have a higher radio-biological effectiveness (RBE) than protons, allowing better control of radio-resistant tumour cells. Indeed, it has been observed that a beam of carbon ions produces, along its track, a great number of clustered, unrepairable 'double strand breaks' on the DNA of the traversed cells. This is related to the six-times-larger charge and the associated 25-times-larger ionization density compared to protons of the same range.

## 2 The first proposals

Linac technology is widely used for radiotherapy. In the world, more than 15,000 electron linacs are used by radiation oncologists, representing about 50% of existing accelerators with energies larger than 1 MeV. The typical energy of the electron beam used for the production of X-rays is in the range 6–20 MeV. Most of the electron-radiotherapy linacs are normal conducting structures working at a frequency of 3 GHz. They are powered by a single power source (magnetron or klystron) and they are typically short and light enough to be mounted on a rotating support, allowing treatment from different angles.

The use of linacs for hadron therapy was proposed for the first time in the late 1980s. Of course, the energy needed for protons is much larger, so proton-therapy linacs cannot be as compact as radiotherapy ones. On the other hand, the main advantage of using a linac for hadron therapy is the possibility of energy modulation, obtained by switching off some units.

### 2.1 The all-linac proposal

In 1991, the first proposal of using a linac for proton therapy was published [2]. The proposed design consisted of a sequence of linear accelerator structures able to accelerate protons up to 250 MeV. A radio frequency quadrupole (RFQ) followed by a drift tube linac (DTL) operating at 499.5 MHz were used as injectors to reach an energy of 70 MeV. Then, a 3 GHz Cell-Coupled Linac (CCL) was added to accelerate the beam from 70 to 250 MeV, as shown in Fig. 2. This type of solution based on an RFQ, a DTL, and a CCL has been named the 'all-linac' approach.

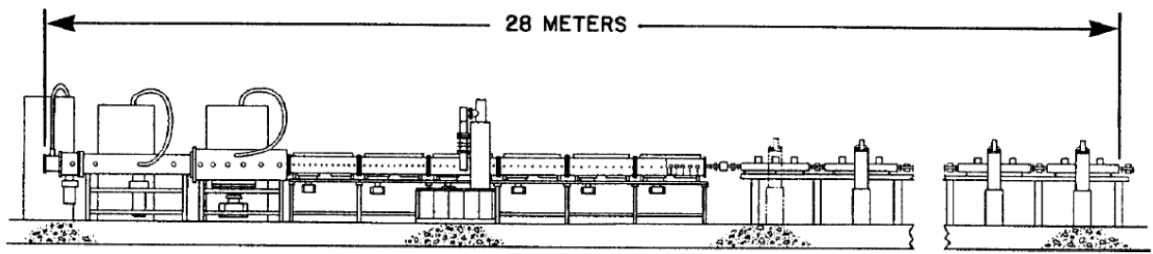

**Fig. 2:** Layout of the PL-250 proton-therapy linac designed in 1991 (taken from [2])

### 2.2 The cyc-linac solution

The so-called 'cyc-linac' solution, first proposed in 1993 [3], is based on the combination of a high-intensity low-energy cyclotron and a high-frequency linac. In this approach, a cyclotron pre-accelerates the particles up to a typical energy of 24–30 MeV. The following high-frequency CCL is used to boost the energy up to 200–230 MeV. In this case, all the accelerating units have the same CCL structure (instead of the three types of linacs needed for the all-linac approach). Furthermore, the cyclotron can be used at night and during the weekends for the production of the radio-isotopes needed for imaging and even for therapy, making the centre a multi-disciplinary environment for physicists, radio-chemists and medical doctors. The first proposed cyc-linac facility is shown in Fig. 3.

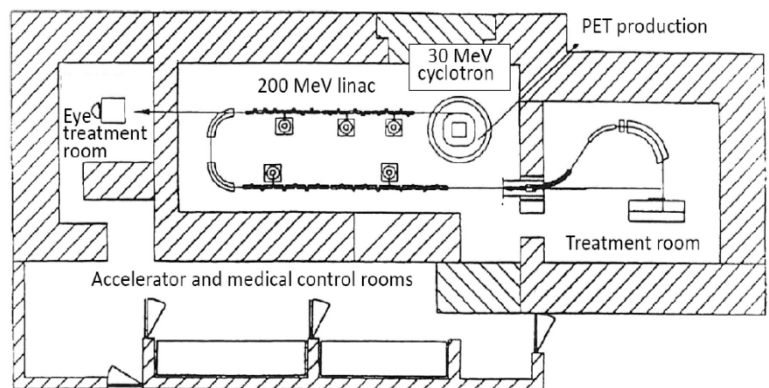

**Fig. 3:** Layout of the first cyc-linac proposal. This type of facility can host both proton-therapy treatment and radio-pharmacy production.

### 2.3 The rationale for proton and carbon-ion tumour therapy with linacs

Up to now, only cyclotrons or synchrotrons have been used for hadron therapy. Most of the proton centres in operation are equipped with cyclotrons while, for carbon therapy, only synchrotrons are used.

Generally, cyclotrons used in proton therapy can reach an energy of 230–250 MeV, are quite compact (with a typical diameter of about 2.5 m), and are continuous wave (CW) machines, so that the beam—bunched by the radio-frequency (RF) cavities at 50–100 MHz—is always present during a patient irradiation. On the other hand, the beam energy can only be varied by passive means, i.e. with motors that introduce absorbers of various thicknesses into the beam. With this method, the time needed for an energy change is of the order of 50–100 ms.

Synchrotrons are more complex and larger than cyclotrons (with a typical diameter of 7–8 m for proton machines and 20–25 m for carbon-ion machines). The beam is extracted in spills of a few seconds with a time separation of 1–2 s. The energy can be varied actively, without needing movable absorbers, by adjusting the number of turns in the machine, but each energy change needs to wait for a new spill. Only very recently, a novel fast extraction technique has been implemented at HIMAC, allowing for multi-energy extraction within one spill [4].

In a linac running at 200 Hz and composed of a large number of accelerating units (typically 10–12) singly powered by independently controlled klystrons, the final beam energy can be varied continuously from pulse to pulse, i.e. every 5 ms, by adjusting the amplitude and/or the phase of the klystron drive signals [5]. The intensity of the beam can also be adjusted on a pulse-to-pulse basis by acting on the source. The particular time structure of a linac pulsed beam is shown in Fig. 4. The number of particles per pulse N can be adjusted from pulse to pulse between the maximum $N_m$ and $N_m/10$, while the energy E can be changed between the minimum energy $E_{min}$ (typically 70 MeV for protons) and $E_{max}$ (230 MeV for protons).

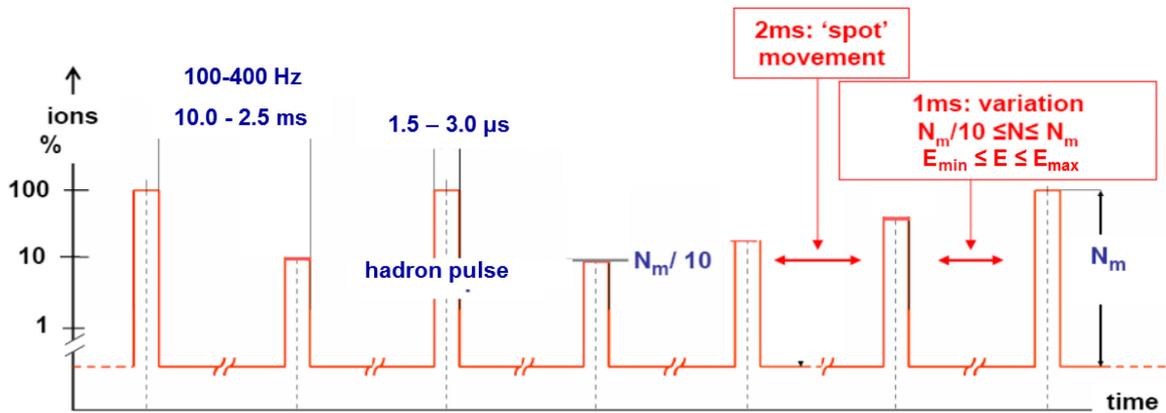

**Fig. 4:** Typical time structure of a high-frequency pulsed linac for hadron therapy. Both energy and intensity of the beam can be varied actively in a few milliseconds.

The 'active' (i.e. obtained by electronic means without needing absorbers) energy and intensity modulation is a unique feature of the linacs discussed in this lecture and makes possible the implementation of the active spot scanning technique with tumour multi-painting, which is considered the best possible way for treating moving organs [6].

### 2.4    Another possible application: linacs for proton imaging

The concept of a linac booster can also be applied to a different application than therapy. Indeed, a proton beam of 350–400 MeV could be used for imaging purposes in so-called proton radiography. In order to fully exploit the benefits of proton therapy, precise imaging of the tumour is needed. In fact, one of the biggest issues is the precise outline of the target volume and the surrounding healthy tissues. Furthermore, the treatment plan for proton beams is typically calculated on images taken with X-rays, and one of the critical issues is the exact conversion from CT (Computed Tomography) numbers to proton-stopping power. At present, such problems are mitigated by the addition of margins to the treatment target volume.

The typical energy of a proton-therapy machine is 230–250 MeV, but in order to have protons traversing the body of the patient (like X-rays used for radiography and CT) a minimum energy of 350–380 MeV is needed. A high-frequency linac coupled to the typical cyclotron used in commercial proton-therapy centres could provide a valuable option to boost the energy of the beam by about 100 MeV, reaching energies of interest for proton imaging. Such types of linac booster were studied by TERA Foundation in collaboration with PSI [7]. A 5 m long linac could be installed along the transfer line going to the last treatment room of cyclotron-based centres, and could be used as a booster when needed or by-passed during normal treatment operation.

## 3    The present

Typical frequencies used for proton linacs in high-energy physics laboratories are 350 or 700 MHz. This is mostly due to the fact that large currents are typically needed. In proton therapy, the requirement for beam current is much less demanding and smaller bore apertures can be used. For this reason, higher frequencies can be used. Furthermore, the availability of S-band power sources already developed by industry for radiotherapy machines has driven the choice of the frequency of the first proton-therapy linac designs. The proof of principle of the use of a 3 GHz linac for acceleration of protons was achieved in 2002 with the LInac BOoster (LIBO) prototype. Since then, several studies and projects with the aim of developing and realizing high-frequency linacs for proton therapy have started.

### 3.1 The LIBO prototype

A 3 GHz Side-Coupled Linac (SCL) unit called LIBO has been designed and built under the leadership of Mario Weiss (CERN) by TERA Foundation in collaboration with CERN and the INFN sections of Milan and Naples. The accelerating unit, shown in Fig. 5, was made of four tanks coupled together by three 'bridge couplers'. In each tank, 13 accelerating cells allow an accelerating gradient of 16 MV/m to be achieved. In the space between each tank, small Permanent Magnet Quadrupoles (PMQs) are installed to create a FODO lattice that focuses the narrow beam through the accelerating structure.

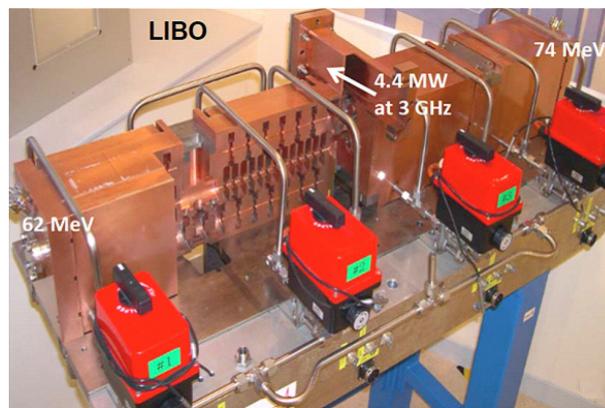

**Fig. 5:** The LIBO-module prototype was the first 3 GHz linac to accelerate protons [8]

The first unit of LIBO has accelerated protons from 62 to 74 MeV at the same 3 GHz frequency of electron linacs. The LIBO project gave birth to other high-frequency linac designs for proton therapy.

### 3.2 The project IMPLART by ENEA

The ENEA group, led by Luigi Picardi and Concetta Ronsivalle, has proposed and built a Side-Coupled Drift Tube Linac (SCDTL) which is better suited than a SCL (because of the larger shunt impedance as discussed in the lecture on 'Accelerating Structures' in these proceedings) to accelerate protons from a few MeV to 40–70 MeV. The SCDTL is made of DTL cells coupled together by coupling cells placed off axis (side-coupled cells) and is designed to work at 3 GHz. The SCDTL structure installed in the ENEA laboratories in Frascati (Rome, Italy) is shown in Fig. 6.

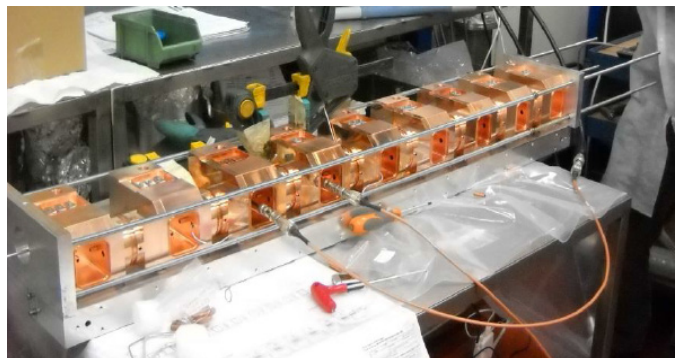

**Fig. 6:** The first SCDTL module tested at CECOM (Guidonia, Italy) before the installation in the ENEA laboratories in Frascati.

This is part of the Intensity Modulated Proton Linear Accelerator for Radio Therapy (IMPLART) which is expected to accelerate protons up to 150 MeV, and after its technical validation at ENEA will be transferred to IFO hospitals in Rome. The all-linac design envisages a commercial RFQ–DTL system working at 425 MHz used as the injector, followed by four SCDTL modules bringing the proton energy up to 35 MeV, and completed by four CCL units to boost the energy up to 150 MeV. In the Frascati ENEA laboratories, the first SCDTL module has been successfully tested up to 11.7 MeV with a proton beam. This is the first time that an SCDTL at such high frequency has been used for proton acceleration [9].

### 3.3 LIGHT by A.D.A.M.

The CERN spin-off company A.D.A.M. (Application of Detector and Accelerators to Medicine), founded in 2007 with the aim of industrializing novel detectors and high-frequency linacs for medical applications, has developed in the past few years a linac for proton therapy, based on the TERA design of LIBO, called LIGHT (Linac for Image Guided Hadron Therapy). The LIGHT design (Fig. 7) is composed of three linear accelerating sections: an RFQ up to 5 MeV, an SCDTL up to 37.5 MeV and a CCL up to 230 MeV.

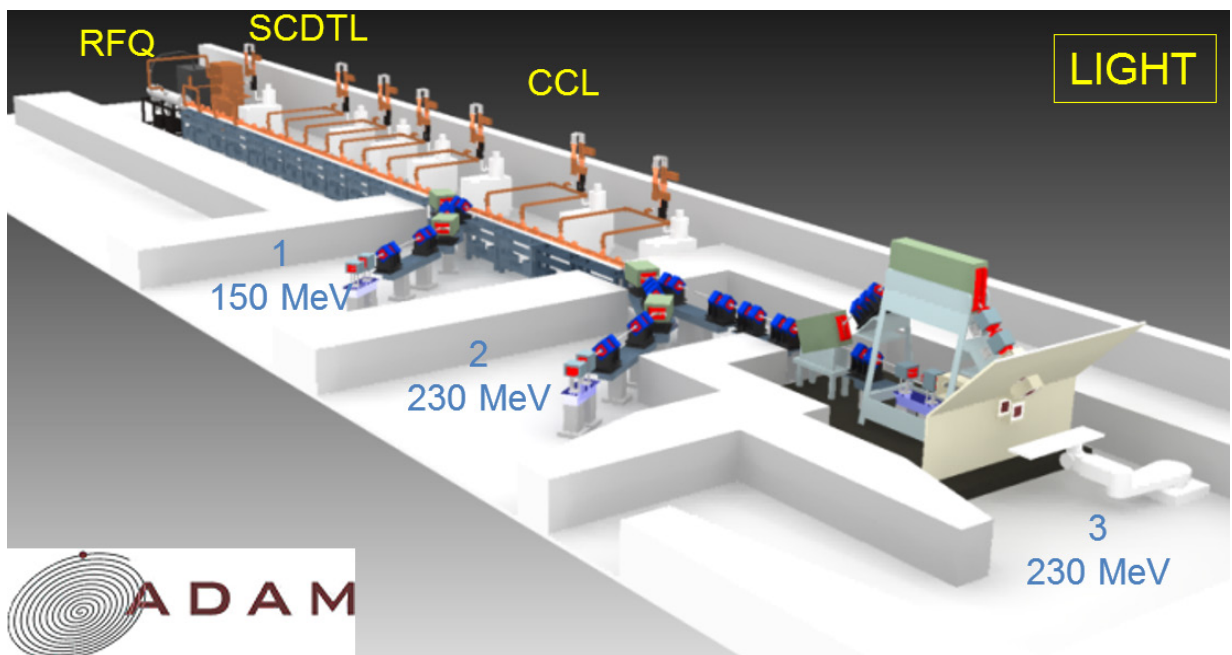

**Fig. 7:** The all-linac design, proposed by A.D.A.M., for LIGHT

The 5 MeV RFQ is based on a novel design at 750 MHz, made by CERN [10]. This is almost double the maximum frequency typically used for this type of machine. A very compact modular solution of only 2 m has been found. A prototype is now being built and is expected to be tested with a beam in the spring of 2016. The second section is based on the 3 GHz SCDTL design by ENEA. The use of the fourth sub-harmonic of 3 GHz for the RFQ makes the matching between the first two sections easier and allows for a compact design. Finally, a 3 GHz CCL section is added to reach 230 MeV. The design of the accelerating units of LIGHT (Fig. 8) has been improved compared to the LIBO prototype, in particular, by reducing the size of the bridge couplers and allowing for an open space between each tank, where the PMQs used to focus the beam can easily be placed.

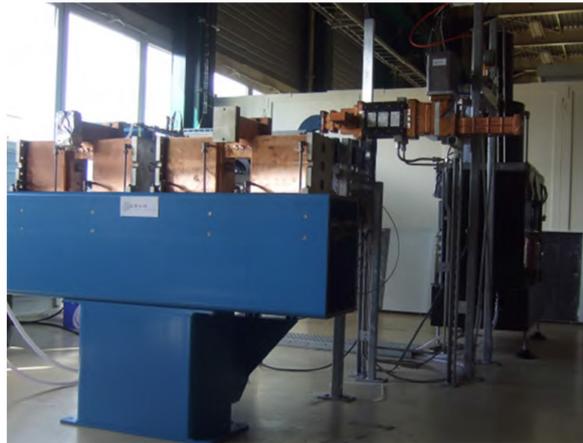

**Fig. 8:** Prototype of the LIGHT first unit built by A.D.A.M.

The first prototype of LIGHT, up to 90 MeV, will be assembled by A.D.A.M. on CERN premises in 2016 and, after an acceleration test, will be mounted in a hospital, followed by the last units needed to accelerate the protons from 90 to 230 MeV.

### 3.4 The novel cyc-linac TERA design

Based on the experience developed after the design and test of LIBO, TERA Foundation has improved the cyc-linac scheme and proposed a novel design for PERLA (Protontherapy and Exotic Nuclei from Linked Accelerators), shown in Fig. 9.

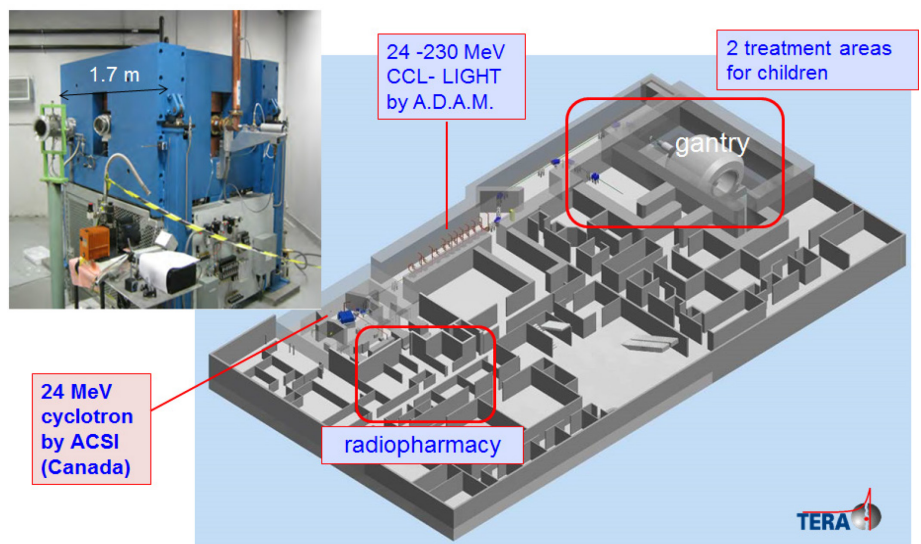

**Fig. 9:** Design of the PERLA accelerator. In this complex, radio-pharmacy production and therapy are combined in the same centre.

The design, based on the cyc-linac concept, combines a 24 MeV cyclotron with a 3 GHz linac, allowing both radio-pharmacy production and proton therapy in the same complex. A preliminary characterization of the beam properties of the TR24 cyclotron produced by ACSI (Canada) has been performed to study the matching line between the two machines [11].

### 3.5 ACLIP design by INFN

After the test of LIBO, the INFN sections of Naples, Milan and Bari worked on a new design to improve the efficiency of SCLs in the energy range from 30 to 62 MeV. A prototype of the first module of ACLIP was built in 2007 and, at the end of 2008, underwent high-power tests in the UK on the premises of the e2V company (Fig. 10). A 4 MW magnetron was used to power the module that was conditioned up to 5.4 µs pulses at 120 Hz [12]. In 2009, a test with a beam was also performed at LNS in Catania, confirming the possibility of using 3 GHz SCL structures for protons, even at 30 MeV.

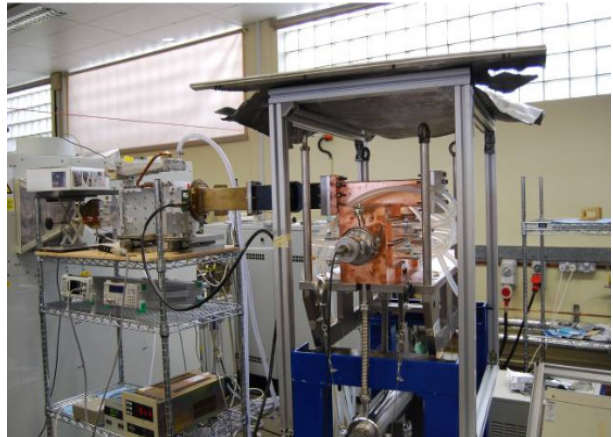

**Fig. 10:** The ACLIP module high-power RF test set

## 4  The future

The main issues that make hadron-therapy linacs more difficult to build and more expensive than typical electron-radiotherapy linacs are the fact that: (i) we want to accelerate hadrons (much heavier than electrons) and (ii) these need to be to an energy of at least 200–230 MeV for protons or 400–430 MeV/u for carbon ions (compared to the maximum of 20 MeV for electrons).

The approximately 45 proton-therapy centres running in the world are 'multi-room' facilities, in the sense that one accelerator typically feeds three treatment rooms. This approach makes good use of the accelerator, but requires long displacements of many patients because the facility serves more than 5 million people. Many experts are convinced that 'single-room' facilities linked to a big hospital and serving 1.5 million people will have a place in the future development of hadron therapy [13].

The radio-biological properties of carbon ions make them suitable for the treatment of radio-resistant tumours and extremely encouraging results have been obtained with more than 10,000 patients treated. Despite this fact, at present there are only eight carbon-ion centres treating patients in the world, and most of them are in Japan [14]. An accelerator for carbon-ion therapy is larger and more expensive than for proton therapy, due to the higher energies needed (400 MeV/u compared to 200 MeV) and the larger beam rigidity (almost a factor of three larger). The use of high-efficiency linacs, with a reduced power consumption and a high-duty cycle, can be envisaged for future dual proton and carbon-ion machines.

### 4.1  Studies for the future: high-gradient hadron structures

The accelerating gradient used in electron linacs is of the order of 20 MeV/ 1 m = 20 MV/m. With the same gradient inside the accelerating structures, the effective accelerating gradient in a structure for low-energy protons—taking into account the transit-time factor (roughly 0.85) and the synchronous phase of the RF field (typically −15°)—would be 16 MV/m. The active length of a 230 MeV proton linac would then be 230 MeV / 16 MV/m = 14.5 m. On top of this, it is important to recall that hadron linacs need focusing elements, whose length, as a rule of thumb, is about 40% of the active length. The actual distance needed for a 230 MeV proton linac—not considering extra space for beam matching or the fact that in the initial acceleration stage, with an RFQ, the gradient is much lower—would then be: 14.5 m × 1.4 = 20 m. This rough estimate gives an idea of the difference in size of proton linacs compared to radiotherapy electron linacs. Of course, for other types of ions, the length becomes a factor of two larger, due to the typical charge to mass ratio $Z/A=1/2$ for the accelerated particles.

In order to make linacs more compact, a larger accelerating gradient is needed. However, an increase in gradient can be limited by several considerations.

First of all, since hadrons are much heavier than electrons, at the energies of interest for therapy, the speed of the particles are typically in the range 0.1 to 0.6 $c$. To increase the acceleration efficiency for such low $\beta$ values, 'nose cones' are typically added along the axis. This results in a better focalization of the field in the accelerating gaps and thus an increase of the transit-time factor. On the other hand, the addition of nose cones enhances the value of the surface electric field $E_s$ (Fig. 11). This is why, in hadron linac structures, the ratio of surface to average field $E_s/E_0$ typically takes values of 4–5 (while for electron linacs $E_s/E_0$ is around 2). In other words, for the same accelerating gradient, the electric surface field in hadron-linac structures is two-times larger. But high surface fields are related to increased probability of breakdowns. Secondly, the power dissipated in the structures scales with the square of the accelerating gradient. A higher gradient means more RF power from the power sources and more heat load that needs to be cooled in order to keep constant the resonant frequency of the structures, and therefore, increased costs.

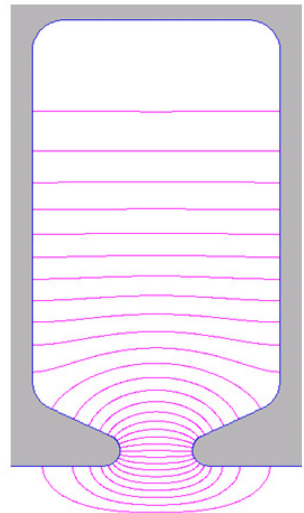

**Fig. 11:** Typical inner shape and electric field lines profile of an accelerating cavity in a CCL hadron linac. The maximum surface electric field is close to the point of the 'nose'.

The CLIC study at CERN has developed a novel accelerating scheme for an electron–positron collider. In the past years, an extensive test program demonstrated the feasibility of 100 MV/m acceleration in normal conducting 12 GHz RF structures [15]. TERA Foundation, in collaboration with the CLIC RF structure development group led by Walter Wuensch, has performed extensive studies on high-gradient accelerating structures for applications in hadron therapy [16]. Even though the aim and the geometry of the TERA and CLIC structures are different, both share the same operational limits in terms of maximum surface electric field $E_{max}$ (~200 MV/m) and of maximum break-down rate (BDR) expressed in breakdown per pulse (bpp) over a certain length (~$10^{-6}$ bpp/m). The first limit is strictly related to the cell geometry, which defines the ratio between peak surface electric field and average accelerating gradient $E_{max} / E_0$. The second limit, concerning linac boosters, corresponds to one breakdown every two treatment fractions (of about 2–3 minutes) at a maximum repetition rate of 200 Hz for a 20 m long linac, which is considered to be acceptable for medical applications.

Two single-cell standing-wave accelerating structures, one at 3.0 GHz and one at 5.7 GHz, have been designed, built and high-power tested [17]. Measurements of the performance in terms of BDR were conducted and compared with the results of single-cell standing-wave X-band accelerating cavities tested at SLAC (Fig. 12).

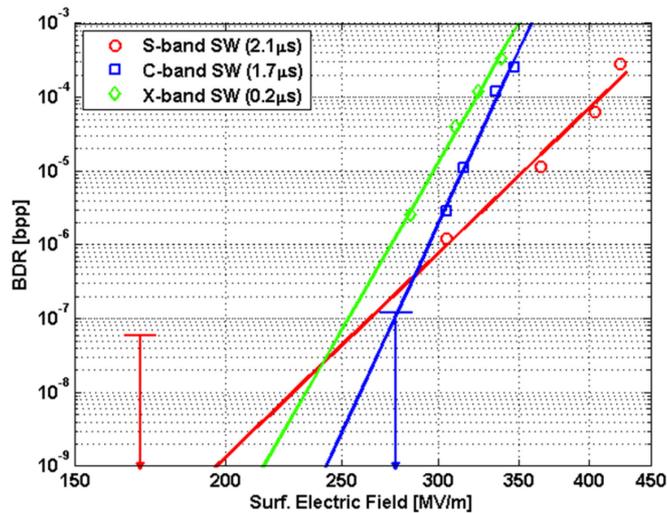

**Fig. 12:** Comparison of results of BDR measurements performed at different frequencies [18]

From these results, it has been concluded that the surface fields in normal conducting structures at 3.0 and 5.7 GHz can be pushed to more than 200 MV/m with a probability of breakdown per cell of less than $10^{-8}$. But even more importantly, the experimental results show that the maximum modified Poynting vector $S_c$—introduced in [19]—describes the BDR measurements in the 3–12 GHz frequency range better than the maximum surface electric field $E_s$.

### 4.2 High-gradient linac for single-room facilities

The project TULIP (Turning LInac for Protontherapy), patented by TERA, envisages a linac, mounted on a rotating gantry, used as a booster for protons previously accelerated by a cyclotron [20]. In the tanks, the maximum average gradient is $E_0$ = 30 MV/m. As is shown in Fig. 13, the RF power transmission is made possible by high-power rotating joints developed in collaboration with the CLIC group.

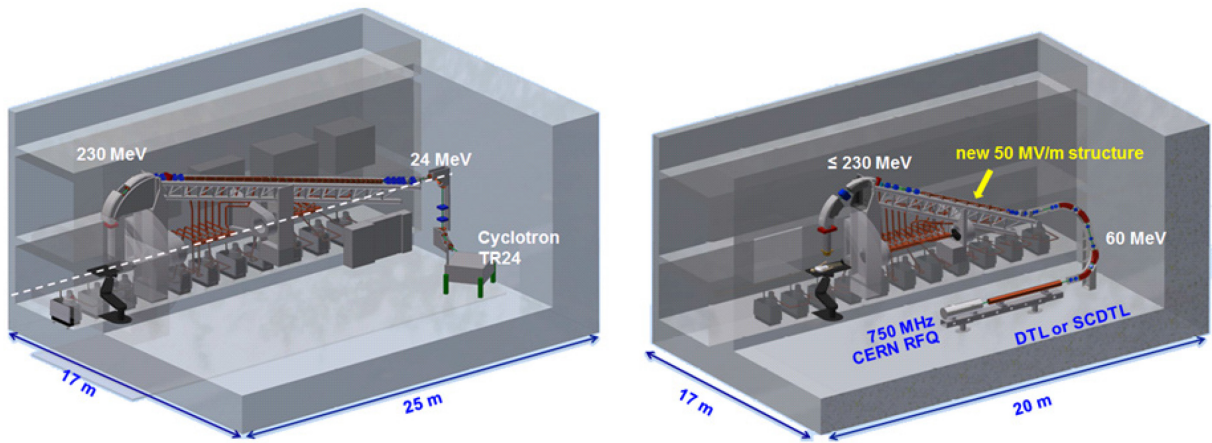

**Fig. 13:** Two possible linac-based solutions for the proton single-room facility TULIP

After several studies, a novel design—based on the optimization of the modified Poynting vector, and using the high-frequency RFQ designed by CERN, followed by an SCDTL as an injector—has been proposed. In this design, average gradients $E_0$ of 50 MV/m in the structures have been considered in order to obtain a more compact solution (as is shown in Fig. 13 (right)).

## 4.3 High-efficiency design for dual proton and carbon-ion facilities

When considering the design of a linac for carbon-ion therapy, the power efficiency of the whole system becomes a challenge. The intrinsic properties of carbon ions are such that:

– the 'spot' size is transversely two-times smaller than the proton spot size (5 mm instead of 10 mm, if the pencil beam has a FWHM= 4 mm) because of the sharper lateral fall-off. In order to keep the treatment time short enough (a few minutes), a higher repetition rate of 300–500 Hz rather than 100–200 Hz is needed;

– concerning acceleration, each proton carries a 'useless' neutron ($Z/A = 1/2$), which means that for 100 MeV/u acceleration, a voltage of 200 MV is needed.

As a consequence, much more average and peak power from the RF sources is required. This is crucial for a carbon-ion facility based on a linac, such as the CABOTO (CArbon BOoster for Therapy in Oncology) design proposed and patented by TERA Foundation [21]. In Figs. 14 and 15, two schematic layouts of CABOTO are presented. The scheme of Fig. 14 is based on a cyc-linac solution. Several designs have been studied for the injection in a linac [22]. As an example, the cyclotron can be similar to the one built by VECC in Kolkata [23], which accelerates light $Z/A = 1/2$ ions up to 70 MeV/u. The following linac—33 m long and pulsed at 300–400 Hz—can boost fully stripped carbon ions from 70 MeV/u to 400 MeV/u. With these figures, the voltage gain of the particles is very large: (400–70) x 2 = 660 MV. The use of very high gradients to reduce the total length (as the ones foreseen for TULIP) would bring the overall power consumption to unacceptable levels (more than 2 MW). This is why the linacs shown in Fig. 14 are designed for a gradient in the structures of 30 MV/m.

Recent developments have enabled significant improvement of the klystron efficiency. Multi-Beam Klystrons (MBKs) at 3 GHz, producing 7 MW RF peak power in pulses of 5 μs at 500 Hz with an efficiency of better than 60% are now available. Similar klystrons, which are the object of a CERN tender launched by the CLIC group, need only 60 kV so that the modulators are oil-free and small. This item will soon be tested at CERN.

With this new type of RF source, the 32 CCL units of CABOTO running at a duty cycle of 1–1.5 $10^{-3}$ (300–400 Hz with 3.5 μs RF pulses) could consume 700 kW. In a full linac configuration, as shown in Fig. 15, with an RFQ, SCDTL and CLL, the maximum average power consumption would be 800 kW. Considering extra ancillaries and power supplies for beam transfer lines, the objective is to run the accelerator complex with a plug-power not larger than 1.2 MW. As a final remark, it has to be underlined that this is the highest power consumption related to the accelerators. While RFQ and SCDTL represent a fixed power consumption, the CCL contribution (which is more than 85%) strongly depends on the depth of the treated tumours.

Apart from the smaller power consumption compared to synchrotrons (1.2 MW at maximum compared to 2.5–3 MW for synchrotrons), the footprint of a dual proton–carbon-ion linac complex can also be smaller than a synchrotron one. In the layouts of Fig. 14, the three treatment rooms and the distribution of the close-by rooms are copied from CNAO in Pavia. The footprints of the accelerators and the power supplies of the two versions of CABOTO are 750 m$^2$ and 1000 m$^2$, to be compared with the 1600 m$^2$ of CNAO.

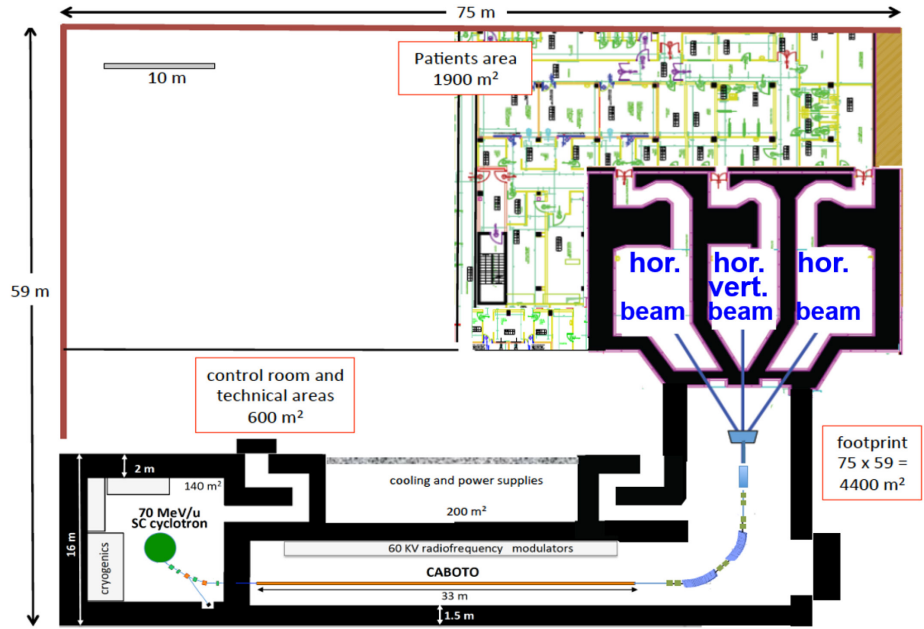

**Fig. 14:** Preliminary layout of a cyc-linac carbon-ion facility based on high-efficiency design

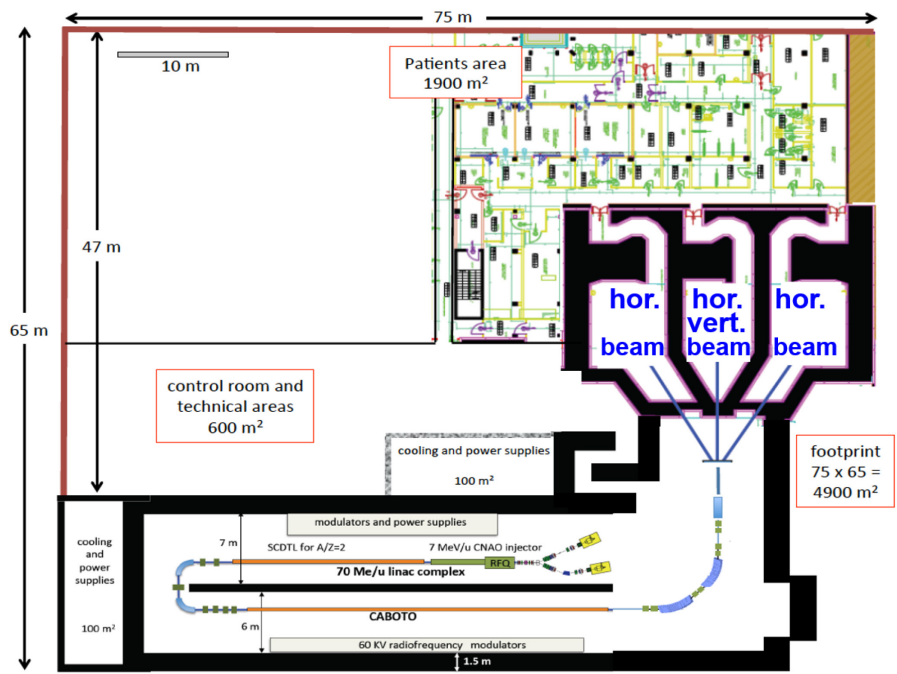

**Fig. 15:** Preliminary layout of an all linac carbon-ion facility based on high-efficiency design

# 5 Summary

High-frequency RF linacs can produce hadron beams that are well suited to treating moving organs with the multi-painting spot scanning technique. The field is now moving from the first prototypes and RF tests to full-scale commercially built linacs.

At present, several groups are working on construction projects. Low-velocity SCDTL and high-velocity CCL accelerating structures have been built and tested by ENEA, TERA, and INFN. At CERN, a new 750 MHz RFQ is being built with the support of the CERN medical application office. The CERN spin-off company A.D.A.M. is building, at CERN, an all-linac facility that will be the first commercial item of this type and will be transferred to a hospital to treat patients.

Future challenges include high-gradient and high-efficiency structures. TERA and the CERN CLIC group have been collaborating for several years on this subject, with the support of the Knowledge Transfer group of CERN. In future, this will lead to TULIP, a compact proton linac rotating around the patient, and to CABOTO, a high-efficiency linac for the therapy of deep-seated radio-resistant tumours with carbon ions.

**Acknowledgement**

I would like to thank Prof. Ugo Amaldi for being the motor of many of the developments described in this lecture and for his contagious passion for physics and health. His comments and the discussions about linacs for medical applications have always been stimulating and enlightening.